\documentclass[oldversion]{aa}

\usepackage{graphicx}
\usepackage{natbib}
\usepackage[varg]{txfonts}

\bibpunct{(}{)}{;}{a}{}{,}

\usepackage{txfonts}

\begin{document}

   \title{The warm absorber in NCG 5548:}

   \subtitle{The lean years}

   \author{R.G. Detmers\inst{1}
          \and
          J.S. Kaastra\inst{1}\inst{,2}
          \and 
          E. Costantini\inst{1}
          \and
          I.M. McHardy\inst{3}
          \and
          F. Verbunt\inst{2}}

   \offprints{R.G.Detmers}

   \institute{SRON Netherlands Institute for Space Research, Sorbonnelaan 2, 3584 CA Utrecht, The Netherlands \email{r.g.detmers@sron.nl}
    \and
    Astronomical Institute, University of Utrecht, Postbus 80000, 3508 TA Utrecht, The Netherlands
    \and
    School of Physics and Astronomy, The University, Southampton SO17 1BJ, UK}     

   \date{}

 
  \abstract{We study the variability of the warm absorber and the gas responsible for the emission lines in the Seyfert 1 galaxy NGC 5548, in order to constrain the location and physical properties of these components. Using X-ray spectra taken with the \textit{Chandra}$-$LETGS in 2002 and 2005, we study variability in the ionic column densities and line intensities. We find a lower \ion{O}{vii} forbidden emission line flux in 2005, while the Fe K$\alpha$ line flux stays constant. The warm absorber is less ionized in 2005, allowing us to constrain its location to within 7 pc of the central source. Using both the observed variability and the limit on the FWHM of the \ion{O}{vii} f line, we have constrained the location of the narrow line region to a distance of 1 pc from the central source. The apparent lack of variability of the Fe K$ \alpha$ line flux does not allow for a unique explanation.}

   
  \maketitle
%

\section{Introduction}			\label{intro}

Active Galactic Nuclei (AGN) have been studied for decades since their discovery over 60 years ago.
Mass loss from the nucleus has been known for a long time to exist in the form of radio jets \citep[see e.g.][]{Peterson97} or in outflows showing broad absorption lines \citep{Weymann81}.
It is only recently however that X-ray and UV observations have shown the presence of outflows in a majority of the moderately luminous Seyfert galaxies \citep{Crenshaw99b}.
These outflows are important as they probe the inner regions of AGN and during the lifetime of an AGN they can constitute a significant mass loss \citep{Blustin05}. They can provide us with a better understanding of the accretion process onto the supermassive black hole and of the enrichment of the intergalactic medium \citep{Hamann99}. The physical structure of these outflows and their connection to the other AGN components, such as the supermassive black hole, the accretion disk and the broad and narrow line region (BLR, NLR), is not understood.
In order to learn more about these outflows and the other components of the AGN, variability studies are very important. These allow one to see how the warm absorber and the emission lines respond to a variable flux from the central source. This puts constraints on the locations of the gas in which they originate, and provides us with information on their physical state.

A good object for the study of these processes is the Seyfert 1 galaxy NGC 5548. It is a relatively nearby AGN \citep[$z$ = 0.01676,][]{Crenshaw99} and has a relatively high X-ray brightness.
Previous UV and X-ray observations have shown the presence of a warm absorber in NGC 5548 \citep{Steenbrugge05,Crenshaw03}. From the kinematics it was concluded that the UV and X-ray absorption lines are part of the same outflow.
By comparing a \textit{Chandra}$-$HETGS and LETGS observation of NGC 5548 in 2002 with an earlier one in 1999, \citet{Steenbrugge05} conclude that there is no evidence for long-term variability, except for the \ion{O}{v} absorption line. 
 
We present the results of a followup \textit{Chandra}$-$LETGS observation of NGC 5548 taken in April 2005, which we compare with previous \textit{Chandra}$-$LETGS and XMM-\textit{Newton} RGS observations \citep{Kaastra02b,Steenbrugge03,Steenbrugge05}.
In all previously mentioned high-resolution observations, the variability in the continuum flux has been at most a factor of 2 and no significant variations have been detected in the emission lines. During our new observation the source was at a very low flux level which lasted at least from March 1 till May 10, as determined from \textit{Swift} observations \citep{Goad06}, optical observations \citep{Bentz07}, and this LETGS observation. During this time the continuum flux was a factor five lower than in 2002. This allows us to analyze the source for the first time at a very low flux level, and determine the changes in the spectral features, such as the warm absorber, narrow emission lines, and the iron Fe K$\alpha$ line.
The \textit{Rossi X-ray Timing Explorer (RXTE)} monitoring data of NGC 5548 from 1996 till 2007 are also presented in order to constrain the long-term variability much more accurately than only using the spectral observations, which have large gaps between them.
We discuss the observation and data reduction in Section \ref{data}, Section \ref{spectral} is used to describe the spectral analysis and the variability of the spectral components.
We discuss our results in Section \ref{discussion} and present our conclusions in Section \ref{conclusions}.  

\section{Data reduction}				\label{data}
   
NGC 5548 was observed for 141 ks with the \textit{Chandra} LETGS \citep{Brinkman00} in April 2005. The data were reduced as described by \citet{Kaastra02b}. In summary, the data were reduced with the standard CXC pipeline up until the level 1.5 event files. For the steps leading to the final event file, level 2, we followed an independent procedure, which is described in the aforementioned article.
The resolution of the LETGS is 0.05 $\AA$ and its band spans a wavelength range of 1 $-$ 180 $\AA$. Because of the low signal to noise ratio and the Galactic absorption towards NGC 5548, we ignore the data above 80 $\AA$ and below 1 $\AA$. The spectrum was analyzed with the SPEX software package \citep{Kaastra96}\footnote{See also http://www.sron.nl/spex}. We use $H_{\mathrm{0}}$ = 70 km $\mathrm{s^{-1}}$ $\mathrm{Mpc^{-1}}$, $\Omega_{\mathrm{m}}$ = 0.7 and $\Omega_{\Lambda}$ = 0.3. Because of the low count rate we calculate the errors using C-statistics \citep[see XSPEC online manual\footnote{http://heasarc.gsfc.nasa.gov/docs/xanadu/xspec/manual/\\ XSappendixCash.html}, we use the form as given by Castor. See also][]{Cash79}. All errors given are 1$\sigma$ errors, computed for $\Delta C$ = 1.
The RTXE data were reduced as described in \citet{Mchardy04}.    

\section{Spectral Features}   	\label{spectral}

\begin{figure}[tbp]
   \includegraphics[angle= -90,width=9cm]{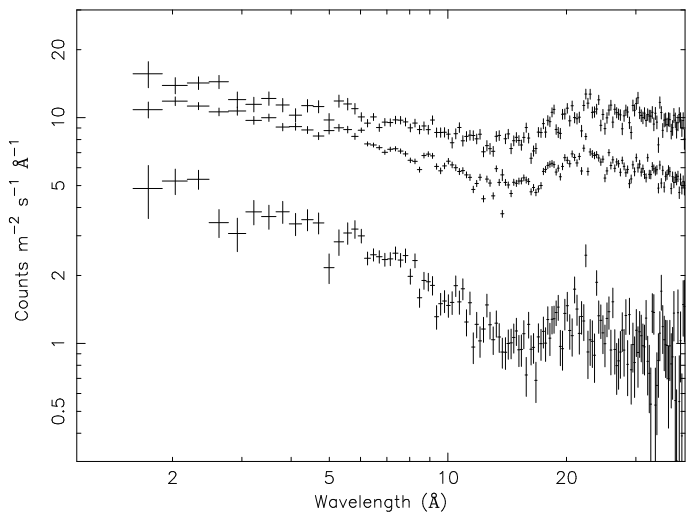}
   \caption{\label{fig:specs}
        The rebinned spectrum of NGC 5548 as taken with the LETGS instrument in 1999 (upper) 2002 (middle) and 2005 (lower).
             }
\end{figure}

Fig.~\ref{fig:specs} shows the spectra from the 1999, 2002 and 2005 observations.
The flux levels are much lower in the most recent data, with a 2 $-$ 10 keV continuum flux of 1.47 $\times$ $10^{-14}$ W $\mathrm{m^{-2}}$. 
To make sure that the spectral lines are analyzed against an adequate local continuum, we fit the continuum with a spline model. A spline model allows us to model the continuum between two boundaries $b_{\mathrm{1}}$ and $b_{\mathrm{2}}$ with a cubic spline. The limits are 1 $\AA$ and 80 $\AA$ in our case. The continuum flux $y_{\mathrm{i}}$ is then given on 159 evenly spaced grid points $x_{\mathrm{i}}$ between $b_{\mathrm{1}}$ and $b_{\mathrm{2}}$ with a spacing of 0.5 $\AA$ between the grid points. The continuum model between the grid points is then determined by cubic spline interpolation. The $y$ values of the grid points are determined by spectral fitting. Using a spline model means we do not assume an a priori shape of the continuum, such as for example a powerlaw and blackbody continuum.
We apply the redshift of 0.01676 for NGC 5548 and correct for Galactic extinction, with the \ion{H}{i} column density fixed to 1.65 $\times$ $10^{24}$ $\mathrm{m^{-2}}$ \citep{Nandra93}. We add various absorption components and narrow emission features to this continuum (see Sect 3.1 $-$ 3.3). The best fit (including all the features) has $C$ = 3228 for 2833 degrees of freedom.

\subsection{Warm Absorber}

In order to model the warm absorber we use the \textit{slab} model in SPEX \citep{Kaastra02a}, which calculates the transmission of a thin, irradiated slab of matter with a set of adjustable ionic column densities. We model the optical depth of the absorption lines with Voigt profiles. Our models consist of 3 slab components, with the ionic column densities set free and with the outflow velocities $v$ fixed to the previous LETGS study \citep{Steenbrugge05}. Simultaneous UV observations of NGC 5548 \citep{Crenshaw03} show five different warm absorber outflow velocities, but due to lower spectral resolution in the X-ray, only three outflow velocities could be distinguished by \citet{Steenbrugge05}, namely $-$1040, $-$530 and $-$160 km $\mathrm{s^{-1}}$.

\begin{figure}[tbp]
   \includegraphics[angle= -90,width=9cm]{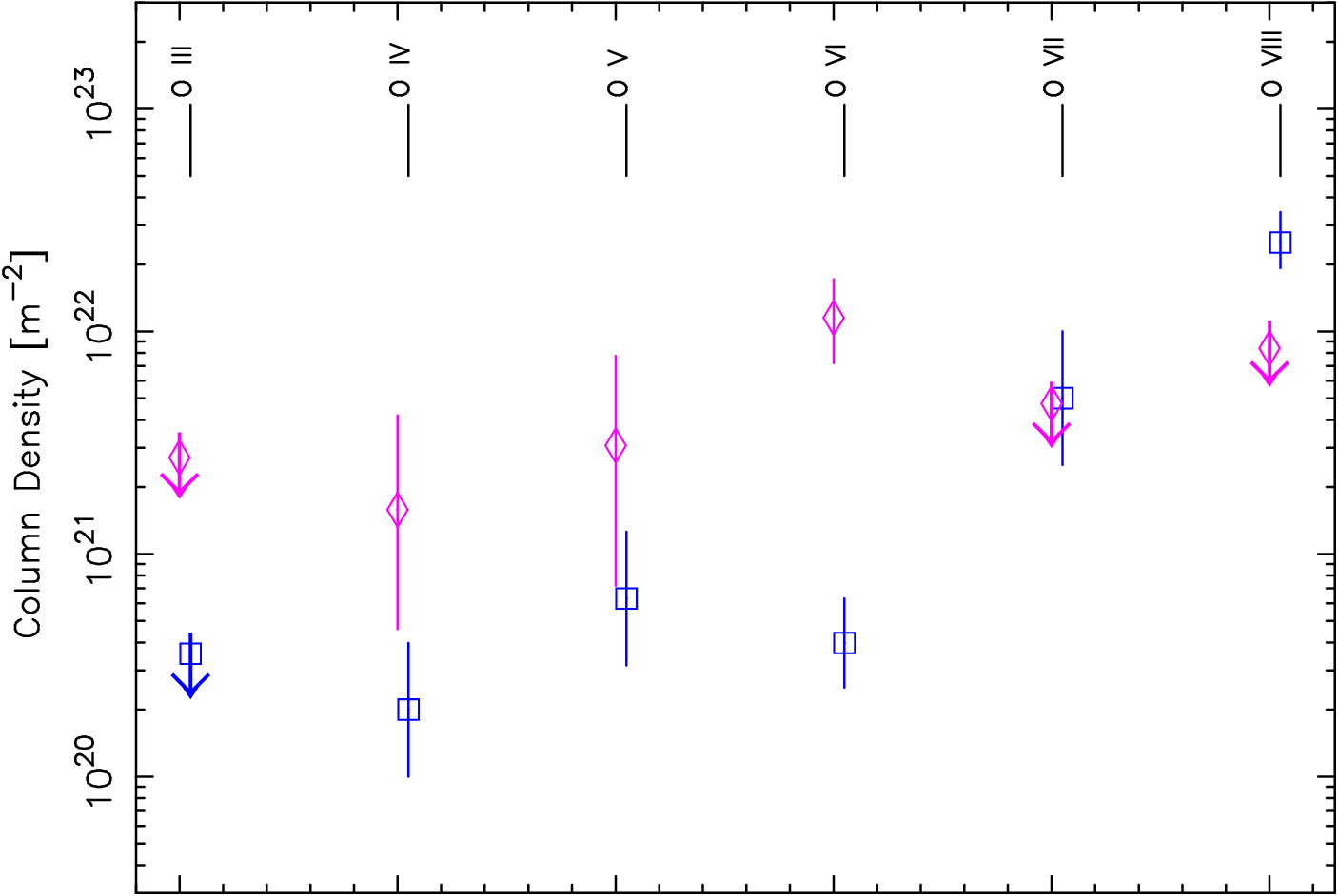}
    \includegraphics[angle= -90,width=9cm]{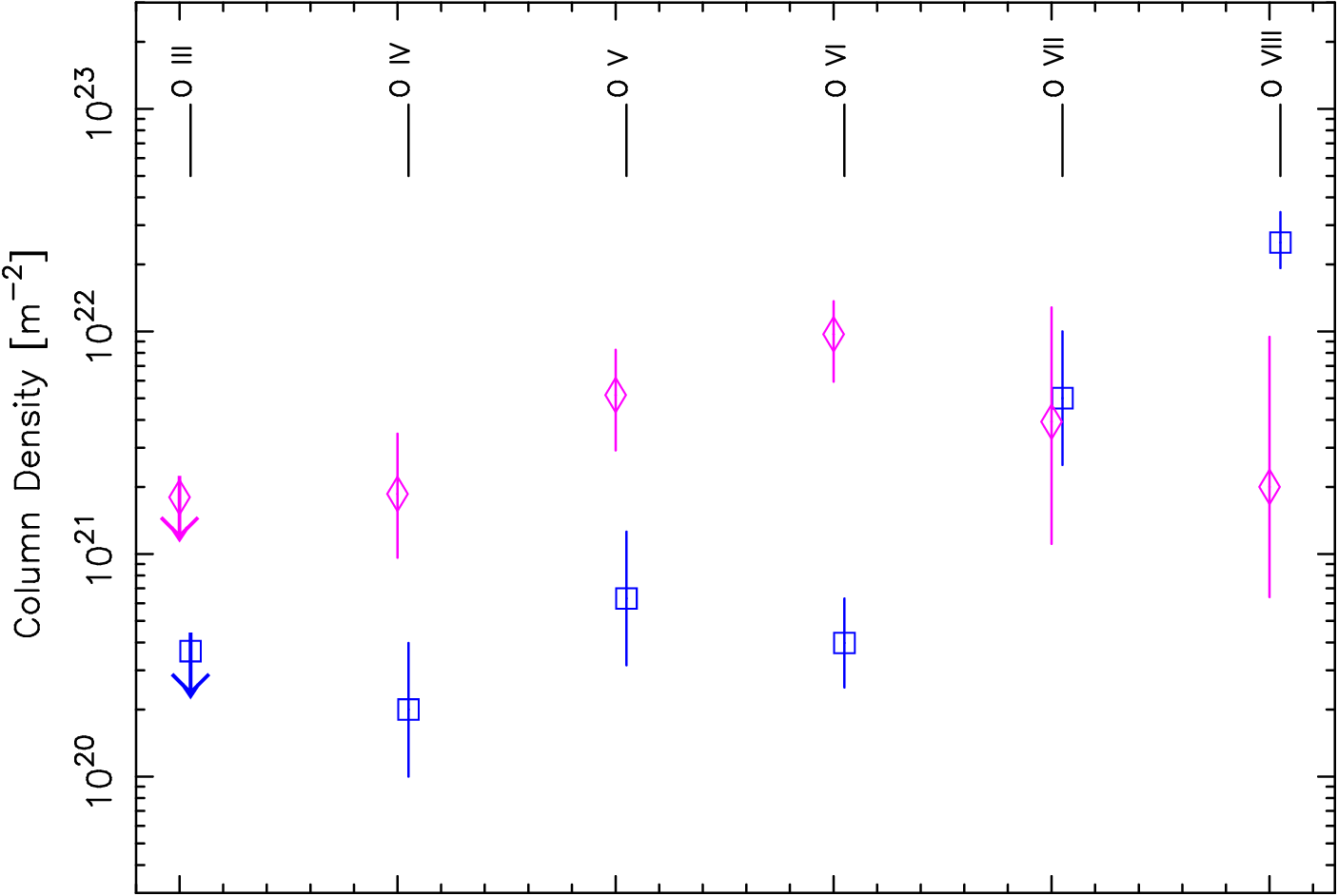}
    \caption{\label{fig:wa}
        Total oxygen column densities as measured with the LETGS instrument in 2002 (rectangles) and 2005 (diamonds). 1$\sigma$ errors are shown as well, except for ions which only have a measured upper limit on the total column density (arrows). The upper panel shows the total column densities with the velocity broadening $\sigma$ set free, the lower for $\sigma$ fixed to the values of \citet{Steenbrugge05}.  }
\end{figure}

A change in the optical depth of an absorption line can be due to a change in the velocity broadening $\sigma$ or a change in the ionic column density $N_{\mathrm{ion}}$. Alternatvely a change in the covering factor of the absorbing gas can also explain the variations in the absorption lines. We assume here a covering factor of unity for the absorbing gas (see Sect \ref{discussion} for discussion).
To be certain that the changes we observe in the absorption lines are indeed due to column density changes and not only due to a possible change in $\sigma$, we performed two fits, one with $\sigma$ set free, the other with $\sigma$ fixed to the values reported in \citet{Steenbrugge05}. 
Fig.~\ref{fig:wa} shows the 2002 and 2005 total column densities for both models. The \ion{O}{viii} column density has decreased, while the \ion{O}{iv} $-$ \ion{O}{vi} column densities have increased. The total column densities are the same for both models.
In Table \ref{column05} we show the 2005 oxygen column densities for both models.

\begin{table}
\caption{Oxygen column densities for the 2005 observation. The logarithms of the column densities are listed in $\mathrm{m^{-2}}$. The velocity broadening $\sigma$ and the outflow velocity $v$ are listed in $\mathrm{km}$ $\mathrm{s^{-1}}$. The lower table is for the same slab model, but with the velocity broadening $\sigma$ fixed to the vales used in \citet{Steenbrugge05}.   }            
\label{column05}      
\begin{tabular}{l c c c c}        
\hline\hline                 

$v$ & -1040 & -530 & -160 & total \\
$\sigma$ & 127 $\pm$ 28 & 62 $\pm$ 21 & 120 $\pm$ 34  \\
Ion             &                  &           &                  &    \\
\hline 
 
\ion{O}{iii} & $<$ 20.5       & $<$ 20.6& $<$ 21.2       & $<$ 21.4         \\      
\ion{O}{iv}  & 20.8 $\pm$ 0.5& $<$ 20.9& 21.0 $\pm$ 0.5& 21.2 $\pm$ 0.5       \\
\ion{O}{v}   & 20.9 $\pm$ 0.6& $<$ 21.3& 21.4 $\pm$ 0.5& 21.5 $\pm$ 0.5      \\
\ion{O}{vi}  & 22.1 $\pm$ 0.2& $<$ 20.9& $<$ 21.4       & 22.1 $\pm$ 0.2      \\
\ion{O}{vii} & $<$ 21.0 & $<$ 20.4& 21.1 $\pm$ 0.5& $<$ 21.6      \\
\ion{O}{viii}& $<$ 21.1 & $<$ 21.7& 21.2 $\pm$ 0.5& $<$ 21.9 \\
\hline\hline                 

$\sigma$ & 40 & 100 & 90  \\

\hline                        

\ion{O}{iii}   & $<$ 20.4 & $<$ 20.4 & $<$ 21.1                         & $<$ 21.2       \\
\ion{O}{iv}   & 20.6 $\pm$ 0.6 & $<$ 21.0 & 21.1 $\pm$ 0.4          & 21.3 $\pm$ 0.3       \\
\ion{O}{v}    & 21.3 $\pm$ 0.4 & 20.5 $\pm$ 0.7& 21.4 $\pm$ 0.3& 21.7 $\pm$ 0.2      \\
\ion{O}{vi}   & 21.9 $\pm$ 0.2 & $<$ 21.3 & 21.1 $\pm$ 0.4          & 22.0 $\pm$ 0.2      \\
\ion{O}{vii}  & 21.5 $\pm$ 0.9& $<$ 20.6& 21.0 $\pm$ 0.5            & 21.6 $\pm$ 0.5      \\
\ion{O}{viii} & $<$21.5 & $<$ 21.5& 21.1 $\pm$ 0.9                       & 21.3 $\pm$ 0.6      \\

\hline                                   
\end{tabular}
\end{table}

\subsection{Narrow emission lines}
The \ion{O}{vii} forbidden line at 22.101 $\AA$ is clearly visible in the spectrum of the 2005 observation. We also searched for the \ion{Ne}{ix} forbidden line at 13.699 $\AA$, since that line is also detected in earlier observations. Table \ref{narrow} shows the strengths of both lines in the new spectrum and in earlier observations. We measure a flux of 0.35 $\pm$ 0.06 ph $\mathrm{m^{-2}}$ $\mathrm{s^{-1}}$ for the \ion{O}{vii} forbidden emission line and we detect a blueshift of -250 $\pm$ 70 $\mathrm{km}$ $\mathrm{s^{-1}}$ for the \ion{O}{vii} f line. For the \ion{Ne}{ix} forbidden emission line we measure an upper limit of 0.04 ph $\mathrm{m^{-2}}$ $\mathrm{s^{-1}}$, if we assume that the \ion{Ne}{ix} line has the same blueshift as the \ion{O}{vii} f line.
Earlier observations as noted in Section \ref{spectral} have reported blueshifts of $-$150 $\pm$ 70 km $\mathrm{s^{-1}}$ \citep[HEG]{Steenbrugge05} and $-$70 $\pm$ 100 km $\mathrm{s^{-1}}$ \citep{Kaastra02b}. The LETGS has a wavelength scale uncertainty of 0.01 $\AA$, or 140 km $\mathrm{s^{-1}}$ at the \ion{O}{vii} f emission line. Taking this and the error on the outflow velocity into account, the blueshift we detect is negligible, but it is however consistent with the HEG measurement \citep{Steenbrugge05}.
The decrease in flux with respect to previous observations is consistent for both lines, although we only detect an upper limit for the \ion{Ne}{ix} f line. 

\begin{table}
\caption{The unabsorbed flux in ph $\mathrm{m^{-2}}$ $\mathrm{s^{-1}}$ for the \ion{O}{vii} and \ion{Ne}{ix} forbidden emission lines for observations taken in 2000, 2001 (RGS, extracted from the public archive, errors include uncertainty due to continuum level) and in 2005. For comparison we also give values from earlier publications for observations taken in 1999 \citep[LETGS,][]{Kaastra02b}, 2000 \citep[MEG,][]{Kaastra02b} and 2002 \citep[LETGS,][]{Steenbrugge05}. All errors are calculated with 68 \% confidence. }             
\label{narrow}      
\centering                          
\begin{tabular}{c c c c}        
\hline\hline                 

Year (month) & Instrument & \ion{O}{vii} & \ion{Ne}{ix} \\    

\hline                        

 1999 (December) &LETGS& 0.81 $\pm$ 0.13 & 0.25 $\pm$ 0.07  \\
 2000 (February) &MEG& 0.82 $\pm$ 0.18 & 0.09 $\pm$ 0.03  \\
 2000 (December) &RGS& 0.65 $\pm$ 0.18 & 0.14 $\pm$ 0.07 \\
 2001 (July) &RGS& 0.70 $\pm$ 0.16 & 0.22 $\pm$ 0.07 \\
 2002 (January) &LETGS& 0.88 $\pm$ 0.08 & 0.14 $\pm$ 0.04 \\      
 2005 (April) &LETGS& 0.35 $\pm$ 0.06 & $<$ 0.04 \\  

\hline
\end{tabular}
\end{table}

\subsection{The narrow Fe K$\alpha$ line}
\begin{figure}[tbp]
   \includegraphics[angle= -90,width=9cm]{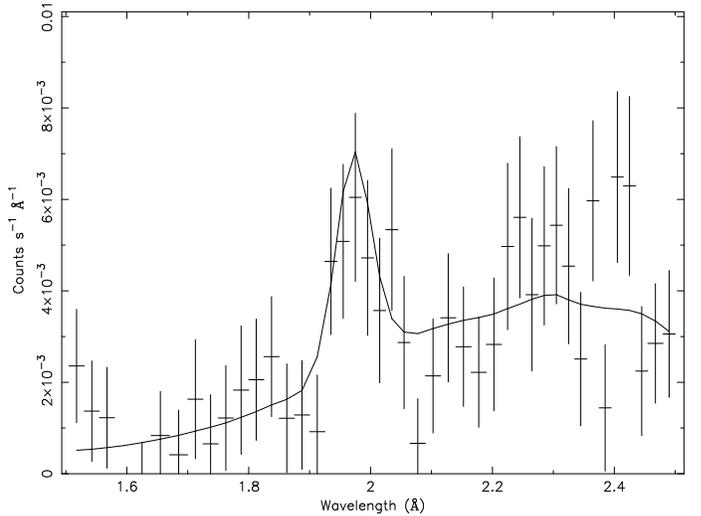}
   \caption{\label{fig:fek}
   The Fe K$\alpha$ emission line fit for the 2005 observation.                }
\end{figure}

The low continuum flux facilitates the detection of a clear Fe K$\alpha$ emission line at 6.39 $\pm$ 0.03 keV in the spectrum of our LETGS data (Fig.~\ref{fig:fek}). Using a Gaussian fit, we measure a flux of 0.55 $\pm$ 0.18 $\mathrm{ph}$ $\mathrm{m^{-2}}$ $\mathrm{s^{-1}}$, which agrees with previous values from \citet{Yaqoob01}, \citet{Pounds03} and \citet{Steenbrugge05} as shown in Table \ref{fek}. The equivalent width of the line is 420 $\pm$ 140 eV, much larger than measured in those previous observations.
There is no evidence for a relativistically broadened Fe K$\alpha$ emission line profile in our spectrum.
We measure a FHWM of 7300 $\pm$ 5100 km $\mathrm{s^{-1}}$, which is also in agreement with previous values.

\begin{table}  
\caption{Fe K$\alpha$ line parameters from the HEG spectrum \citep[as reported by][]{Yaqoob01}, XMM-\textit{Newton} EPIC spectrum \citep{Pounds03}, HEG spectrum \citep{Steenbrugge05} and present LETGS spectrum (2005). The FWHM is given in km $\mathrm{ s^{-1}}$, the flux in ph $\mathrm{m^{-2}}$ $\mathrm{s^{-1}}$.  All errors are calculated with 68 \% confidence.  }             
\label{fek}      
\centering                          
\begin{tabular}{l@{\,}c@{\,}c@{\,}c@{\,}c}        
\hline\hline                 

 & HEG(2000)  & EPIC(2001) & HEG(2002) & LETGS(2005) \\    

\hline                        
 
 E (keV)&6.40$\pm$0.02&6.39$\pm$0.01&6.39$\pm$0.01&6.39$\pm$0.03 \\      
 EW (eV)&130$\pm$30&60$\pm$9&47$\pm$11&420$\pm$140 \\
 FWHM&4500$\pm$1600& 6500 $\pm$ 1300& 1700 $\pm$ 1100&7300$\pm$5100 \\
 Flux&0.36$\pm$0.10&0.38$\pm$0.06&0.24$\pm$0.06&0.55$\pm$0.18\\

\hline                                   
\end{tabular}
\end{table} 

\section{Discussion}				\label{discussion}

\subsection{The location of the warm absorber}

\begin{figure}[tbp]
   \includegraphics[angle= -90,width=9cm]{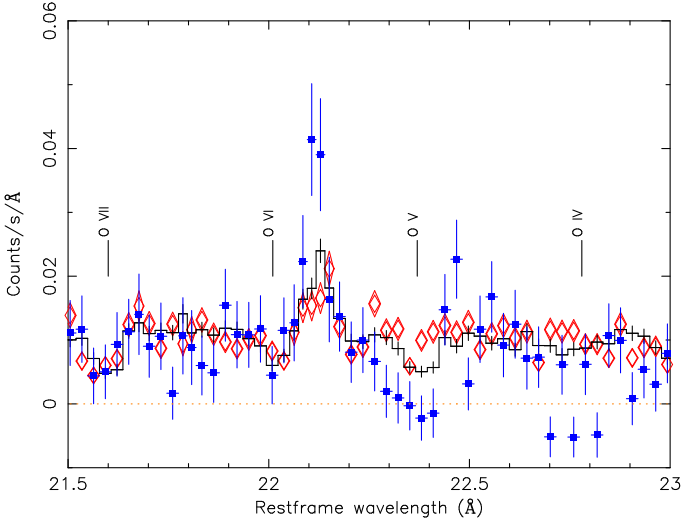}
    \caption{\label{fig:spectra}
        Comparison of the 1999 (diamonds), 2002 (solid line) and 2005 (rectangles) spectra, near the forbidden \ion{O}{vii} emission line at 22.101 $\AA$. The deep \ion{O}{v} absorption line is visible at 22.354 $\AA$.  The 1999 and 2005 spectra have been scaled to the 2002 flux.  }
\end{figure}

Fig.~\ref{fig:wa} shows that the \ion{O}{viii} column density has decreased, while the \ion{O}{iv} $-$ \ion{O}{vi} column densities have increased between 2002 and 2005, both for $\sigma$ fixed and $\sigma$ set free. Some errors on the oxygen column densities in the 2005 spectrum are large, due to a combination of low statistics and saturation of some of the absorption lines. 
As shown in Fig. \ref{fig:spectra} there are additional indications that the column density of the lower oxygen ions has increased from 2002 to 2005. The \ion{O}{v} absorption line at 22.354 $\AA$ is indeed much stronger and broader than in either 1999 or 2002. The equivalent width of the absorption line is 140 $\pm$ 30 m$\AA$, compared to 44 $\pm$ 6 m$\AA$ in 2002 and 22 $\pm$ 9 m$\AA$ in 1999.
Allowing the velocity width $\sigma$ to vary, the column densities of the individual velocity components do change, but the total column densities do not change significantly (see Table \ref{column05}). This is a further indication that the observed changes in the absorption lines are indeed due to variations in the column density and not only due to a possible change in $\sigma$. 

In principle this still leaves open the possibility of a varying covering factor of the x-ray absorbing gas. As is shown in UV spectral observations of NGC 5548 \citep{Crenshaw03}, the covering factor for the various components is not unity. In the UV one can resolve the different velocity components of the absorber, while this is very hard, if not impossible to achieve with the resolution of the current X-ray instruments. The size of the X-ray emitting region (10 $R_{g}$ is estimated to be much smaller than the size of the UV emitting region. Also the BLR region is larger for the optical / UV than for the X-rays \citep[see e.g.][]{Costantini07}. This, along with the fact that the column densities for the highly ionized oxygen species have decreased, while column densities of the lowly ionized species have increased, allows us to be quite confident with the assumption of a covering factor of unity for the X-ray absorbing gas. 
\newline
The large increase in \ion{O}{vi} can be explained by recombination of \ion{O}{viii} and \ion{O}{vii}.

The ionization parameter $\xi$ is determined by the ratio of the ionizing flux and the density of the gas \citep{Tarter69}:

\begin{equation}		\label{xi}
\centering
      \xi = \frac{L}{n R^{2}}. 
\end{equation}
Here $L$ is the 1 $-$ 1000 Rydberg luminosity, $n$ the density of the gas and $r$ the distance from the ionizing source. 
In fact, one can only determine the product of the density and distance squared, $n$$R^{2}$, from the luminosity and $\xi$. The luminosity $L$ = $9.0 \times 10^{36}$ W in the 2002 LETGS observation \citep{Steenbrugge05}. 
We will first consider the changes in the \ion{O}{viii} column density to obtain a lower limit for the density $n$. The column density of \ion{O}{viii} in the 2002 LETGS observation is $3.39 \times 10^{22}$ \citep[See model B,][]{Steenbrugge05}. We assume that the value for the ionization parameter is the one where the column density of the ion peaks, so log $\xi$ = 1.65 for \ion{O}{viii}, using the XSTAR code\footnote{http://heasarc.gsfc.nasa.gov/xstar/xstar.html} for photoionization modelling. 
We assume that the changes we detect are due to recombination to determine $n$.
The density $n$ scales inversely with the recombination time $\tau_{\mathrm{rec}}$ of the gas and depends on the recombination rate of the specific ion \citep{Krolik95,Bottorff00}:

\begin{equation}		\label{trec}
\centering
      \tau_{\mathrm{rec}}(X_{i}) = \left({\alpha_{\mathrm{r}}(X_{i})n \left[\frac{f(X_{i+1})}{f(X_{i})} - \frac{\alpha_{\mathrm{r}}(X_{i-1})}{\alpha_{\mathrm{r}}(X_{i})}\right]}\right)^{-1}, 
\end{equation}
where $\alpha_{\mathrm{r}}(X_{i})$ is the recombination rate from ion $X_{i=1}$ to ion $X_{i}$ and $f(X_{i})$ is the fraction of element $X$ in ionization state $i$. 
If $\tau_{\mathrm{rec}}$ is positive, it means that the amount of \ion{O}{viii} increases, by recombination of \ion{O}{ix}. If $\tau_{\mathrm{rec}}$ is negative, then this means that more \ion{O}{viii} is destroyed by recombination to \ion{O}{vii} than that which is created by recombination of \ion{O}{ix}. So the sign of $\tau_{\mathrm{rec}}$ determines, whether \ion{O}{viii} is destroyed or created and the value of $\tau_{\mathrm{rec}}$ shows how long it takes before equilibrium is reached.

As the change in column densities occurred between Jan 21 2002 and April 15 2005, the upper limit on $\tau_{\mathrm{rec}}$ is 1160 days (in the rest frame of the source). The $f(X_{i+1})$ to $f(X_{i})$ for i = \ion{O}{viii} ratio can be determined from the ionization parameter $\xi$. For log $\xi$ = 1.65, the fraction of \ion{O}{ix} to \ion{O}{viii} is 0.48. From $\xi$ we derive a temperature of 86\,000 K. With this value of $T$, $\alpha_{\mathrm{r}}(\ion{O}{vii})$ = 8.65 $\times$ $\mathrm{10^{-18}\, m^{3}\, s^{-1}}$ and $\alpha_{\mathrm{r}}(\ion{O}{viii})$ = 1.31 $\times$ $\mathrm{10^{-17}\, m^{3}\, s^{-1}}$ \citep{Nahar99}. Using these values we find a lower limit on the density of 4.2 $\times$ $\mathrm{10^{9}\, m^{-3}}$.
Using this limit for $n$ and using (\ref{xi}), we get an upper limit of the distance of $R$ $<$ 7 pc. 
If we instead use \ion{O}{v} to determine the distance, using log $\xi$ = -0.2 and $T$ = 19\,000 K, we get an upper limit of $R$ $<$ 19 pc

Another method of determining the location of the warm absorber is to use its column density $N_{H}$ and its thickness $\Delta$$R$, namely $N_{H}$ = $n$ $\times$ $\Delta$$R$. 
We again use the \ion{O}{viii} column density \citep[log $\xi$ =  1.65, $N_{H}$ = 7.6 $\times$ $10^{25}$ $\mathrm{m^{-2}}$ ][]{Steenbrugge05}. Since the gas has responded to the lower flux within 1160 days, the upper limit on $\Delta$$R$ is the light travel distance of 1 pc. 
From this we get a lower limit on $n$ of $2.5 \times 10^{9}\, \mathrm{m^{-3}}$. Again using (\ref{xi}), we get an upper limit of the distance of $R$ $<$ 9 pc.

The main uncertainty causing the large upper limits is of course the large gap in the spectral data between 2002 and 2005. 

\subsection{The variable $\ion{O}{vii}$ f narrow emission line}
The line flux of the \ion{O}{vii} f line has decreased from 0.88 $\pm$ 0.12 to 0.35 $\pm$ 0.06 ph $\mathrm{m^{-2}}$ $\mathrm{s^{-1}}$ between 2002 and 2005 (Table \ref{narrow}). This immediately gives us an upper limit to the distance, namely the light travel distance of 1 pc. 
A lower limit of the distance to NLR gas can be obtained from the velocity dispersion and the mass of the supermassive black hole as given by \citet{Netzer90}:

\begin{equation}		\label{fwhm}
\centering
M = \frac{3 r V_{\mathrm{FWHM}}^{2}}{4 G},
\end{equation}
with $M$ the mass of the supermassive black hole, $V_{\mathrm{FWHM}}$ the FWHM of the line and $r$ the distance of the line emitting region. This assumes an isotropic velocity distribution, which is reasonable, since we can not obtain any detailed line profile information for the NLR region.
Using $M$ = (6.54 $\pm$ 0.26) $\times$ $10^{7}$ $\mathrm{M_{\odot}}$ \citep{Bentz07}, and an upper limit to $V_{\mathrm{FWHM}}$ of 560 km $\mathrm{s^{-1}}$, which we have obtained by using a Gaussian line model for the \ion{O}{vii} f line instead of a delta line, we obtain a lower limit of 1.2 pc for the location of the NLR. 
If instead the NLR would follow the same distribution in velocity as the BLR in AGN \citep{Peterson04}, then the lower limit would be 0.9 pc. A similar limit of $\sim$ 1 pc was also derived by \citet{Kaastra03}.
Although this lower limit can change by a factor of a few, depending on the exact velocity distribution and the upper limit can be in fact smaller as it is based on sparse temporal sampling, both limits used together constrain the location of the NLR to about 1 pc. This is the first time that a well-constrained location has been determined, due to the combination of variability and line width limits.  
Optical measurements \citep{Kraemer98} indicate that the optical narrow line emission comes from a region within 70 pc, with a high ionization component at 1 pc, in agreement with our findings.

We have checked whether $\tau_{\mathrm{rec}}$ is much smaller than $\tau_{\mathrm{lt}}$, so that the total delay is determined only by $\tau_{\mathrm{lt}}$:

\begin{equation}		\label{tdelay}
\centering
      \tau_{\mathrm{delay}} = \tau_{\mathrm{lt}} + \tau_{\mathrm{rec}},
\end{equation}
with $\tau_{\mathrm{delay}}$ the total delay between changes in the continuum flux and changes in the emission line flux, $\tau_{\mathrm{lt}}$ the light travel time through the emitting region and $\tau_{\mathrm{rec}}$ the recombination timescale.
Using (\ref{xi}), with log $\xi$ = 1 (where the \ion{O}{vii} fraction in the gas peaks) and $R$ = 1 pc, we get a lower limit to $n$ of 1 $\times 10^{12}$ $\mathrm{m^{-3}}$. Again from $\xi$ we can determine the temperature for the gas with log $\xi$ = 1, which is 33000 K. The fraction \ion{O}{viii} / \ion{O}{vii} = 0.25 for $\xi$ = 1.
  
Using the \ion{O}{vii} and \ion{O}{vi} recombination rates from \citet[ $\alpha_{\mathrm{r}}(\ion{O}{vii})$ = 1.65 $\times 10^{-17}$ $\mathrm{m^{3} s^{-1}}$ and $\alpha_{\mathrm{r}}(\ion{O}{vi})$ = 0.98  $\times 10^{-17}$ $\mathrm{m^{3} s^{-1}}$]{Nahar99}, the upper limit on $\tau_{\mathrm{rec}}$ is 3 days, much less than the 1160 days limit from variability arguments. This means the total delay time $\tau_{\mathrm{delay}}$ is determined only by the light travel time $\tau_{\mathrm{lt}}$.

\subsection{The origin of the Fe K$\alpha$ line}
Due to the low continuum flux, we are able to detect the Fe K$\alpha$ emission line with the LETGS, even though the LETGS is not optimized for the wavelength range where the line is present, since its spectral resolution is 50 m$\AA$. The interesting result is that the line flux is in agreement with earlier observations, while one would expect a decreased line flux, if the line flux correlates with the continuum flux. Due to the low continuum flux and the constant line flux the equivalent width (EW) of the line is large, 420 $\pm$ 140 eV.

Using both HEG measurements of the Fe K$\alpha$ line (Table \ref{fek}), the weighted average of the FWHM is 2600 $\pm$ 900 km $\mathrm{s^{-1}}$.
Once again using (\ref{fwhm}), the line is formed at 0.06 $\pm$ 0.04 pc from the central source, assuming again a mass of $\sim$ 6.54 $\times$ $10^{7}$ $\mathrm{M_{\odot}}$. At this distance one would expect that the iron line would also have responded to the lower ionizing flux, since the light travel time is much less than 3 years.

There are at least five possible explanations for the constant flux and corresponding larger EW of the emission line. 
The first one which can immediately be ruled out is the light-bending near the black hole \citep{Miniutti04}. The derived distance from the FWHM of the line (0.06 pc, or 19\,000 R$_{\mathrm{g}}$), is much too large for any light-bending to play a role.

The second reason why the iron line is not affected by the lower continuum could be that the line consists of multiple unresolved components originating from a broad range of distances from the central source. If we assume that this is the case, then at least part of the emission could come from a region further away from the central source than 1 pc, for instance the torus structure at a parsec-scale \citep{Antonucci93}. However \citet{Suganuma06} have performed infrared reverberation mapping of NGC 5548 and found a time-lag of 47$-$53 days between the continuum flux in the $V$ band and the IR emission (dust). This would put the dust at $\sim$ 0.04 pc. The dust can of course extent much further outward, so Fe K$\alpha$ emission from reflection on dust further than 1 pc away from the central source cannot be ruled out. However, if the bulk of the Fe K$\alpha$ line would originate from pc distances, the line should be much narrower than what is observed.

Another explanation is that the continuum experienced a peak in flux just before our observation, such that the Fe K$\alpha$ line would have already responded to this increase in continuum flux, but the warm absorber and the narrow emission lines did not. From the LETGS observation alone this cannot be ruled out, but there were also optical observations \citep[March 1 till April 10,][]{Bentz07} of NGC 5548 just before the LETGS observation. They detected a very low continuum and also a weaker H$\beta$ emission line. Since the H$\beta$ line has a time-lag of 6.3 days, any peak in continuum flux should have shown up in the optical observations.
Also the RXTE monitoring does not show a peak in the months prior to our observation.

The fourth possibility is based on the change in continuum flux and the effect it has on the gas which reflects the Fe K$\alpha$ line. \citet{Matt93} have shown that the strength of the Fe K$\alpha$ line depends on the ionization parameter $\xi$ of the reflecting gas. They show that if $\xi$ $\leq$ 100, the line strength increases with decreasing $\xi$. Since the continuum flux has dropped, it is not unreasonable to assume that the ionization state of the gas producing the Fe K$\alpha$ line will be lower. So this means that the line flux could stay constant when the continuum flux drops. 

The final possibility is that while the continuum flux in the 2 $-$ 10 keV band has gone down dramatically, the reflection component above 10 keV which produces the Fe K$\alpha$ line may be much less variable than the continuum. This has been detected recently in several AGN with \textit{Suzaku} \citep{Reeves06}. The reason for the lack of strong variability in the reflection component is still unknown.
Due to the errors on the Fe K$\alpha$ line flux or the instrumental resolution of the LETGS, we cannot distinguish among these explanations.

\subsection{Constraints from RXTE monitoring}

\begin{figure}[tbp]
   \includegraphics[angle= -90,width=9cm]{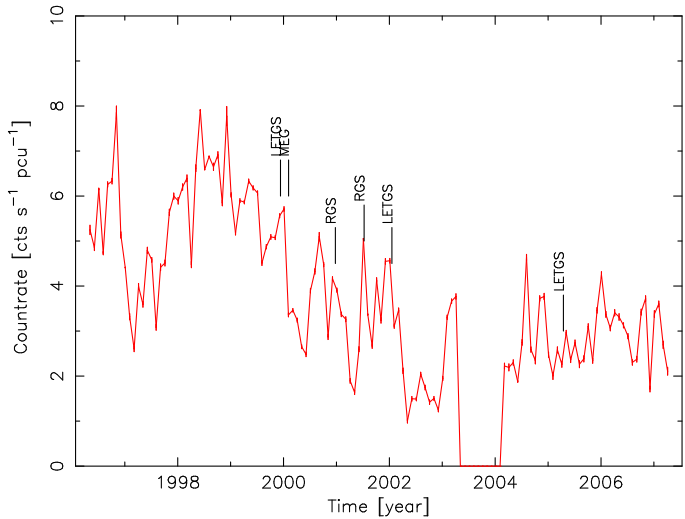}
    \caption{\label{fig:rxte}
    \textit{RXTE} lightcurve of NGC 5548, re-binned to one month intervals, spanning from 1996 till 2007. The count rate shown is per PCU detector of the \textit{RXTE}. All 6 spectral observations (see Table \ref{narrow}) are shown as well. There is a gap in the data from the beginning of 2003 till 2004, when the source was not monitored with \textit{RXTE}. The errors on the countrates are at the 1 \% level.            }
\end{figure}
    
Figure \ref{fig:rxte} shows the long time scale light curve of NGC 5548 as observed with RXTE from 1996 till 2007. The data have been re-binned to a time scale of one month. As can be seen, the variability on a time scale of a few months can be larger than that of longer time scales, i.e. years. There still are long-term changes visible. From 2002 on, the average count rate is 20 $-$ 25 $\%$ lower than that of the earlier years. 
In principle with the information we have on the continuum flux of NGC 5548 between 2002 and 2005, we can further refine our estimate of the upper limit on the recombination time scale for the warm absorber.

For the upper limit to the variability time scale of the warm absorber we currently use an upper limit of 1160 days. Based only upon the two LETGS spectra, this timescale is wildly uncertain. We note that the \textit{RXTE} continuum flux was at approximately the same level in mid-2004 as it was during the 2002 LETGS observation. So if the ionization timescale is short, the ionization state in mid-2004 might be the same as in 2002. In which case a better estimate for the recombination timescale might be 1 year, and even that might be far too long. E.g. there is a peak in the X-ray flux only a few months before our 2005 observation, so an upper limit of a few months might be even a better estimate. If the recombination timescale is 4 months instead of 1160 days, the upper limit on the warm absorber distance would be 2 pc instead. 
As this recombination timescale is so uncertain, and likely to be a serious over-estimate, we therefore conclude that the warm absorber is probably much further in than the 7 pc we derived as an upper limit.
Further studies, taking time-dependent ionization into account, will be able to better model the changes in the warm absorber as it reacts to the continuum variations that \textit{RXTE} has observed.

Concerning the changes in the emission lines, previous observations have shown no observable change in flux for both lines (see Tables \ref{narrow} and \ref{fek}), even though the continuum varied substantially between those observations. This is a clear indication that both emission lines respond to the continuum changes on a long time scale, i.e. years not weeks or months. For the Fe K$\alpha$ line this would mean that any correlation between the continuum flux and the line flux is more complex than a linear relation between the two. 
Possible correlations between the emission lines and continuum flux will be investigated further in a future paper.      

\section{Conclusions} 		\label{conclusions}
This is the first detailed X-ray analysis of NGC 5548 at an extremely low flux level. The warm absorber which has shown little signs of change in earlier observations shows clear signs of recombination to a lower ionized state. We have constrained its location to within 7 pc of the central source, although this is most likely a serious overestimate. The narrow forbidden \ion{O}{vii} line has decreased in flux, which along with the limit on the line width, places its location at $\sim$ 1 pc from the central source. This location also agrees with previous estimates for this source.
Only the Fe K$\alpha$ emission line has not responded linearly to the lower continuum flux, the line flux is consistent with earlier observations. The most likely explanations are a change in the ionization state of the reflecting gas or a constant reflection component which does not follow the 2 $-$ 10 keV continuum.

\begin{acknowledgements}
SRON is supported financially by NWO, the Netherlands Organization for Scientific Research.       
\end{acknowledgements}

\bibliographystyle{aa}
\bibliography{bibfiles}

\end{document}